\documentclass[aps,10pt,twocolumn,preprintnumbers,prd,showpacs,nofootinbib]{revtex4}

\usepackage{amssymb,amsmath}

\usepackage[
    colorlinks,%
    linkcolor=blue,citecolor=red,urlcolor=blue,
]{hyperref}
\usepackage{tikz}
\usepackage{graphicx}
\usepackage{wrapfig}
\usepackage{calc}

\def\barr{\begin{array}}
\def\earr{\end{array}}
\def\ben{\begin{equation}}
\def\een{\end{equation}}
\def\bena{\begin{eqnarray}}
\def\eena{\end{eqnarray}}

\newcommand{\R}{{\mathbb R}}

\newcommand{\M}{\mathcal{M}} 
\renewcommand{\S}{\mathcal{S}} 

\newcommand{\fb}{F\mathcal{M}} 

\newcommand{\gh}{z} 
\newcommand{\hk}{y} 

\newcommand{\ggh}{\mathfrak{z}} 

\newcommand{\sect}{\section}

\def\stackto #1 { \, {\stackrel{#1}{\longrightarrow}}\, }
\def\stackTo #1 { {\stackrel{#1}{\Longrightarrow}} }

\newcommand{\SO}{{\rm SO}}
\newcommand{\so}{\mathfrak{so}}

\newcommand{\ISO}{{\rm ISO}}
\newcommand{\Iso}{\mathfrak{iso}}

\newcommand{\SU}{{\rm SU}}

\newcommand{\g}{\mathfrak{g}}

\newcommand{\half}{\frac{1}{2}}

\newcounter{Ccounter} 
\newenvironment{C-list}{  
\begin{list}{{\rm C\arabic{Ccounter}}.}{\usecounter{Ccounter}}
}{\end{list}}

\newcounter{Cpcounter} 
\newenvironment{C'-list}{  
\begin{list}{{\rm C\arabic{Cpcounter}${}'$}.}{\usecounter{Cpcounter}}
}{\end{list}}

\newcommand{\grqc}[1]{\href{http://arxiv.org/abs/gr-qc/#1}{arXiv:gr-qc/#1}}

\newcommand{\arxiv}[1]{\href{http://arxiv.org/abs/#1/}{arXiv:#1}}

\newcommand{\webpage}[1]{{\color{blue}}\url{#1}{\color{blue}}}

\begin{document}

\title{Contracted Lorentz invariance for gravity with a preferred foliation}
\author{Steffen Gielen}
\email{sgielen@perimeterinstitute.ca}
\affiliation{Riemann Center for Geometry and Physics, Leibniz Universit\"at Hannover, Appelstra\ss e 2, 30167 Hannover, Germany}
\affiliation{Perimeter Institute for Theoretical Physics, 31 Caroline St. N., Waterloo, Ontario N2L 2Y5, Canada} 

\begin{abstract}
In canonical gravity, the choice of a local time direction is not obviously compatible with local Lorentz invariance. One way to address this issue is to view gravity as a gauge theory on {\em observer space}, rather than spacetime. In a Lorentz covariant theory, observer space is the space of unit future-directed timelike vectors; picking such a vector locally breaks the symmetry to a subgroup $\SO(D)$ of $\SO(D,1)$, so that on observer space the local symmetry group is $\SO(D)$. Observer space geometries naturally describe any gravitational theory that only assumes local invariance under $\SO(D)$. Here we construct {\em nonrelativistic} observer spaces for gravity with a fixed foliation, applicable to preferred foliation theories such as Ho\v{r}ava--Lifshitz gravity. Different Ho\v{r}ava--Lifshitz observers at a point are related by a change in the shift vector, leaving the preferred foliation invariant. Gravity is formulated on a nonrelativistic frame bundle with structure group $\ISO(D)$; the shift vector acts as a symmetry breaking field breaking $\ISO(D)$ symmetry spontaneously to $\SO(D)$. A natural torsion-free connection is constructed, including the usual $\SO(D,1)$ connection of relativistic gravity plus terms depending on derivatives of the shift vector. This observer space geometry provides a novel geometric framework for the study of preferred foliation theories.
\end{abstract}

\preprint{pi-cosmo-321}
\pacs{04.50.Kd, 02.40.Ma, 11.15.Ex, 11.30.Cp}
\date{3 May 2013}

\maketitle

\sect{Introduction}

Symmetry is one of the most important guiding principles in physics. Lorentz invariance in particular has become the cornerstone of the modern picture of the fundamental interactions; the standard model of particle physics is built on the global Lorentz symmetry of Minkowski spacetime, while general relativity is based on the equivalence principle which locally implements the relativity of inertial frames, related by Lorentz transformations in special relativity. More concretely, general relativity in $D$ spatial dimensions\footnote{Even though some discussions in this paper are tied to the special case $D=3$, most of it is valid for general $D$.} can be formulated as a theory of a spin connection $\omega$ on a principal $\SO(D,1)$ bundle, the {\em orthonormal frame bundle} $F\M$, together with a soldering form or coframe field $e$ which can be viewed as an isomorphism between $\R^{D,1}$ and each tangent space $T_x\M$, soldering the fibers of an associated vector bundle with fibers $\R^{D,1}$ to spacetime $\M$. One can pull back the 1-forms $\omega$ and $e$ on $F\M$ along a local trivialization of $F\M$ (which exists over any open set in $\M$ that admits a smooth frame) to obtain local 1-forms on spacetime $\M$, also called $\omega$ and $e$. In terms of these variables, the Einstein--Hilbert--Palatini action
\bena
S_{{\rm EHP}}[\omega,e] & = & \frac{1}{16\pi G}\int_\M\epsilon_{abcd}\left(e^a\wedge e^b\wedge R^{cd}[\omega]\right.\nonumber
\\&&-\left.\frac{\Lambda}{6}e^a\wedge e^b\wedge e^c\wedge e^d\right)
\label{akshn}
\eena
defines vacuum general relativity in $3+1$ dimensions with a cosmological constant $\Lambda$ (an analogous formulation exists for all $D\ge 2$).

Vacuum gravity can be formulated purely in terms of the metric, and so introducing a local $\SO(D,1)$ invariance seems unnecessary, but once spinors are coupled, one needs to introduce a coframe and a spin connection. The action (\ref{akshn}) is separately invariant under spacetime diffeomorphisms on $\M$ (bundle automorphisms from the perspective of $F\M$) and local Lorentz transformations; it is independent of the choice of local section used to define the spacetime 1-forms $\omega$ and $e$. 

In canonical gravity one chooses a foliation splitting spacetime into space and time; the diffeomorphisms leaving Eq.~(\ref{akshn}) invariant can be foliation-preserving or foliation-changing. In a general gauge theory, gauge transformations and spacetime symmetries would be unrelated, but in gravity the soldering form maps representations of $\SO(D,1)$ to powers of the tangent and cotangent bundle, which are affected by diffeomorphisms. It is hence not obvious how the $\SO(D,1)$ invariance of Eq.~(\ref{akshn}) is affected by a choice of foliation. In particular, the compatibility of the Ashtekar--Barbero formulation of canonical gravity with Lorentz invariance has been a debated issue in the past. Recently \cite{asht}, a formulation was given that interprets the choice of foliation as specifying a {\em field of observers}, fixing a time direction at each spacetime point. This field of observers spontaneously breaks Lorentz symmetry to $\SO(D)$ in gravity, explaining the appearance of $\SU(2)$ as the gauge group of Ashtekar--Barbero variables. It is then natural to consider {\em observer space}, which includes all local choices of time direction. The orthonormal frame bundle over spacetime becomes an $\SO(D)$ bundle over observer space, and it is there where the different variables in the Ashtekar--Barbero formulation are most naturally defined \cite{obspaper}. 

The idea that gravity might include a preferred foliation in its most fundamental formulation has attracted a lot of interest recently, mainly due to the proposal of Ho\v{r}ava--Lifshitz gravity \cite{horava} that might address the notorious problem of renormalising gravity. In a preferred foliation theory, one could similarly wonder about the fate of local Lorentz invariance; one might expect it to be broken, given that Ho\v{r}ava's basic idea was to incorporate anisotropic scaling between space and time,
\ben
\vec{x}\mapsto \lambda \vec{x}\,,\quad t\mapsto \lambda^z t\,,
\label{aniso}
\een
into the fundamental symmetries of a gravity theory.

In this paper, we will extend the constructions of Ref.~\cite{obspaper} to preferred foliation theories such as the Ho\v{r}ava--Lifshitz theory. We start by considering the dynamical fields defined on an $\SO(D)$ bundle over spacetime $\M$ and reconstruct the associated observer space $O$, so that transformations between observers leave the foliation invariant; these transformations correspond to changing the shift vector. The associated $\SO(D)$ bundle over $O$ then becomes a {\em nonrelativistic frame bundle} over spacetime $\M$, for which the structure group is the Galilei group $\ISO(D)$.\footnote{Of course the Galilei group describes a principle of relativity of inertial frames just as well as the Lorentz group. Here we use ``nonrelativistic'' in the sense of ``not special-relativistic'' as is common in much of the physics literature.} The spin connection on this bundle compatible with a given orthonormal frame, specified by a normal to the foliation and a cotriad in the spatial hypersurfaces, is determined by the Galilean Cartan's equations of structure. Since these differ from their Lorentzian counterparts, the resulting connection is not the one that would arise from the Lorentz-covariant formulation (\ref{akshn}); it includes all the components of the usual spin connection but also has additional components not fixed by the equations of structure.

After explaining the relevance of Cartan geometry to situations of symmetry breaking in gravity, we start by considering an obvious example of a Lorentz-violating theory: Newtonian mechanics, in the geometric formulation initiated by Cartan in Ref.~\cite{cartangeo} and reviewed and extended, e.g., in Ref.~\cite{newtcart}. In this setting we are not able to view the $\SO(D)$ bundle over observer space as a principal bundle over spacetime while preserving the structure of Newton--Cartan gravity; transformations between observers cannot be seen as gauge symmetries of Newton--Cartan gravity. This seems due to the existence in Newton--Cartan gravity of not only {\em absolute time} but also {\em absolute space} and the distinction between spatial geometry and acceleration due to a Newtonian force. The only admissible kinematical group preserving both absolute space and absolute time is the static group $\SO(D)\ltimes\R^{2D+1}$; it fixes a preferred observer field but does not separate spatial curvature and Newtonian acceleration.

For Ho\v{r}ava--Lifshitz gravity, we exhibit a set of transformations relating different observers, which preserves absolute time but not absolute space. These correspond to changes in the shift vector. The corresponding nonrelativistic frame bundle is the natural structure to describe spacetime geometry in preferred foliation theories. It is then natural to require, as in usual general relativity without fermions, the vanishing of torsion of the Cartan connection $A$ on the nonrelativistic frame bundle. This fixes the $\so(D)$ part of $A$ to be the $\so(D)$ part of the usual torsion-free spin connection. The boost part is, however, less constrained than in the standard formulation; some of its components are equal to the usual extrinsic curvature, but there are additional components usually constrained to vanish. We give a possible choice of the additional components compatible with the extended $\ISO(D)$ gauge symmetry on the nonrelativistic frame bundle. $A$ is then torsion-free if and only if the {\em projectability} condition of Ho\v{r}ava--Lifshitz gravity on the lapse holds; this is the condition on $N$ to only depend on time. It is curious that the projectability of the lapse hence arises as a condition on parts of the \mbox{curvature of $A$.}

Our construction offers a novel viewpoint on spacetime geometry with a preferred foliation where Lorentz invariance is not broken but {\em contracted}: on the nonrelativistic frame bundle, there is still a notion of transformations between inertial frames, given not by Lorentz but by Galilei transformations. The shift vector $\vec{N}$ appearing in canonical formulations of gravity can be viewed as a symmetry breaking field that breaks Galilei symmetry locally to a subgroup $\SO(D)$.

\sect{Cartan Geometry and Symmetry Breaking}
\label{cartansec}

It is well-known that there are several ways of extending the gauge group of vacuum general relativity beyond $\SO(D,1)$. After Kibble had realized that the spin connection $\omega$ and the coframe $e$ could be unified into a connection valued in the Lie algebra of the Poincar\'e group $\ISO(D,1)$ \cite{kibble}, an interesting reformulation of general relativity with a cosmological constant which makes this unification idea explicit was given by MacDowell and Mansouri \cite{macman}, who showed that the action
\ben
S_{{\rm MM}}[A] = -\frac{3}{32\pi G\Lambda}\int \epsilon_{abcd4}\left(F^{ab}[A]\wedge F^{cd}[A]\right)
\een
reduces, up to a topological Gauss--Bonnet term, to general relativity with a (positive) cosmological constant in $3+1$ dimensions (\ref{akshn}) if one identifies the $\so(4,1)$ connection $A$ as
\ben
A = \begin{pmatrix} \omega & \sqrt{\frac{\Lambda}{3}}e \\ -\sqrt{\frac{\Lambda}{3}}e & 0 \end{pmatrix}\,,
\label{cartanident}
\een
where we split $\so(4,1)$ into a subalgebra $\so(3,1)$ and a complement $\ggh$. While the MacDowell--Mansouri formulation formally succeeds in unifying the dynamical fields $\omega$ and $e$ into a single object $A$, for which the curvature can be identified with the curvature and torsion associated to $(\omega,e)$, the action breaks the invariance under $\SO(4,1)$ transformations explicitly to a subgroup $\SO(3,1)$. Already the splitting of $\so(4,1)$ used to define the fields $\omega$ and $e$ requires a (noncanonical) choice of $\so(3,1)$ subalgebra, breaking $\SO(4,1)$ covariance. This choice of subalgebra amounts to reducing an $\SO(4,1)$ bundle on which $A$ is initially defined to an $\SO(3,1)$ bundle, to be identified with the orthonormal frame bundle, on which the {\em Cartan connection} $A$ is defined as a 1-form for which the $\so(3,1)$ part is an Ehresmann connection $\omega$ and for which the $\ggh$ part can be viewed as a soldering form. (For the differential geometric foundations of this construction, see Ref.~\cite{sharpe}.)

A more compelling formulation where this symmetry breaking occurs {\em spontaneously} was given by Stelle and West \cite{stellwest}, who introduced an additional field $\gh$ valued in de Sitter space, viewed as the unit hyperboloid in $\R^{4,1}$. At each point in spacetime, the value of $\gh$ determines the splitting of $\so(4,1)$ into a subalgebra $\so(3,1)_\gh$ leaving $\gh$ invariant and a complement $\ggh_\gh$. By allowing transformations of the field $\gh$ as well as the field $A$, one can maintain full $\SO(4,1)$ covariance. The terminology ``spontaneous symmetry breaking'' that we take from Ref.~\cite{stellwest} here refers to the analogy with spontaneous symmetry breaking in particle physics, where a group $G$ is broken to a subgroup $H$ by a field taking values in the coset space $G/H$. In order to have a dynamical mechanism of symmetry breaking, one could think of adding a Higgs-like potential for $z$, as was speculated on in Ref.~\cite{stellwest}. In the rest of this paper, the symmetry breaking fields we consider are {\em observer fields}; a natural class of dynamical observers appear in {\em dust models} for canonical gravity such as in Ref.~\cite{brownku}, where a dynamical dust field is added to gravity to have a local standard of space and time. Our observers, however, are nondynamical gauge fields transforming under the enlarged symmetry group $G$.

Geometrically, the role of the connection $A$ is to specify how a copy of de Sitter space with cosmological constant $\Lambda$ traces out the geometry of the manifold $\M$ as it is rolled along it \cite{macdo}. This defines a {\em Cartan geometry} on spacetime $\M$---a generalization of Lorentzian geometry where the geometry of spacetime is described not by comparing tangent Minkowski spacetimes $\R^{3,1}$ to different spacetime points but by modelling $\M$ infinitesimally on a copy of de Sitter space, the homogeneous space $\SO(4,1)/\SO(3,1)$, which plays a role analogous to the space of symmetry-breaking vacua $G/H$ in particle physics. Cartan geometry is the geometric framework to describe situations of symmetry breaking in gravity \cite{derekproc}.

A similar instance of symmetry breaking occurs in canonical gravity, where one breaks up the unity of spacetime by introducing a foliation of spacetime $\M$ into spatial hypersurfaces $\S_t$, determined by a 1-form $\hat{u}$ normal to these hypersurfaces. The frame field ${\bf e}$ dual to $e$ maps such a 1-form (appropriately normalized) into a spacetime scalar $\hk$ valued in the hyperboloid $H^D \cong \SO(D,1)/\SO(D)$. We can think of this hyperboloid as the space of observers sitting at a point in spacetime, related to one another by Lorentz transformations. The spin connection $\omega$ and the coframe $e$ now split into representations of the subgroup $\SO(D)_\hk$ leaving $\hk$ \mbox{invariant \cite{asht},}
\ben
\omega = \begin{pmatrix} \Omega & K \\ K & 0 \end{pmatrix}\,,\quad e = \begin{pmatrix} E \\ y\,\hat{u} \end{pmatrix}\,.
\label{split1}
\een
Defining a spacetime observer field $u$ by $y=e(u)$, one can then split $\omega$ into a {\em temporal} component proportional to $\hat{u}$ and the {\em spatial} complement; $E$ is spatial by construction, $E(u)=0$. Rewriting the Einstein--Hilbert--Palatini--Holst action (which can only be defined in $D=3$, due to the existence of two linearly independent symmetric invariant bilinear forms on the Lie algebra $\so(3,1)$ \cite{dereksigma}),
\bena
S_{{\rm EHPH}}[\omega,e]& =& \frac{1}{8\pi G}\int\half\epsilon_{abcd}e^a\wedge e^b\wedge R^{cd}[\omega]\nonumber
\\&&+\frac{1}{\gamma}e^a\wedge e^b\wedge R_{ab}[\omega]\,,
\label{holstak}
\eena
using Eq.~(\ref{split1}) and the split of $\Omega$ and $K$ into temporal and spatial parts, one recovers the usual Ashtekar--Barbero formulation of canonical gravity \cite{asht}; the dynamical variables are the cotriad $E$ and the {\em Ashtekar--Barbero connection}, the spatial part of $\Omega+\gamma\,K$ [this sum being understood to be a sum of $\so(3)$-valued 1-forms, using the isomorphism of the adjoint and fundamental representations of $\so(3)$]. Unlike in the usual formulation \cite{holst} where the $\SO(3,1)$ symmetry of Ref.~(\ref{holstak}) is broken explicitly to $\SO(3)$ by imposing ``time gauge'' $e^0_i=0$, here the breaking occurs spontaneously; it is specified by the value of the field $\hk$ at each point in spacetime. The split of spacetime into space and time and the breaking of local Lorentz invariance are intimately tied together by the existence of a soldering form $e$ in gravity.

The work of Ref.~\cite{asht} was not the first attempt at obtaining an $\SO(3,1)$-covariant formalism for Ashtekar--Barbero variables; see, e.g., Ref.~\cite{cianfrani}, using a field $\chi$ which plays a role analogous to the observer field $y$. The advantage of the formalism of Ref.~\cite{asht} is its geometric appeal: $y$ is directly related to the choice of foliation, i.e., to lapse and shift, instead of being an independent additional object.

While this provides a compelling explanation for why the gauge group for Ashtekar variables is the rotational group $\SU(2)$ instead of the Lorentz group, the geometric interpretation in terms of Cartan geometry is clearest if one thinks of $\omega$ and $e$ as parts of a spacetime Cartan connection $A$, defined as in Eq.~(\ref{cartanident}). The natural ingredients for a Cartan connection ${\bf A}$ on {\em space} $\S_t$ are ${\bf\Omega}$, the spatial part of $\Omega$, and the cotriad $E$; for a vanishing cosmological constant, $A$ is an $\ISO(3,1)$ connection, and ${\bf A}$ is accordingly valued in the Galilei algebra $\Iso(3)$,
\ben
{\bf A}=\begin{pmatrix} {\bf\Omega} & E/l \\ 0 & 0 \end{pmatrix}\,,
\label{cartangeo}
\een
where we introduce an arbitrary length scale $l$ for dimensional reasons. The geometry of space $\S_t$ is modelled on ``space'' in the model Minkowski spacetime, defined as the spacelike plane orthogonal to an observer $y$; the Ashtekar--Barbero formulation can be reinterpreted as {\em Cartan geometrodynamics}, a framework of evolving spatial Cartan geometries \cite{asht, obspaper}.

The complete picture hence includes two different types of symmetry breaking: starting with a model Minkowski spacetime with isometry group $\ISO(3,1)$, we can either first pick a spacetime point, breaking the symmetry to $\SO(3,1)$, and then an observer at that point, or we can first pick an observer, breaking to $\ISO(3)$, and then a point in the space relative to this observer. In both cases we have eventually picked an observer in Minkowski spacetime: a point together with a normalized timelike vector at that point. 

Motivated by this observation, we can describe the geometry by going from spacetime to observer space, the space of unit future-directed timelike vectors. Observer space $O$ naturally carries a Cartan connection of the form
\ben
A = \begin{pmatrix} \Omega & K & \vec{e}/l \\ K & 0 & e_0/l \\ 0 & 0 & 0 \end{pmatrix}\,,
\label{cartanobs}
\een
which specifies the geometry of $O$ by modelling it infinitesimally on $\ISO(D,1)/\SO(D)$, the space of observers in Minkowski spacetime. This $A$ is {\em the same} connection that is used to describe spacetime geometry: as we mentioned before, even though in physics one often thinks of $A$ as an $\Iso(D,1)$-valued 1-form on $\M$, such a 1-form is really the pullback of a connection 1-form on the orthonormal frame bundle $\fb$ along some section. Now the orthonormal frame bundle is not only an $\SO(D,1)$ bundle over spacetime $M$ but also an $\SO(D)$ bundle over observer space $O$ (a local observer can be completed to an orthonormal frame by choosing $D$ spacelike vectors). As shown in Ref.~\cite{obspaper}, the conditions that $A$ must satisfy to be a Cartan connection on $\fb\to \M$ imply those for $\fb\to O$.

Fixing a field of observers $u$, as we have done in the above construction, means fixing a section of the bundle $O\to\M$. We can pull back the bundle $\fb\to O$ along this section:
\ben
\begin{picture}(80,75)
\put(0,0){$\M$}\put(15,3){\vector(1,0){50}}\put(70,0){$O$}\put(40,6){$u$}
\put(0,60){$Q_u$}\put(15,63){\vector(1,0){50}}\put(65,60){$F\M$}\put(40,66){$i_u$}
\put(8,55){\vector(0,-1){45}}\put(75,55){\vector(0,-1){45}}
\end{picture}
\een
The pullback $u^*A$ of $A$ to $Q_u\to\M$ then inherits a splitting of $\Iso(D,1)$ into representations of $\SO(D)$ from Eq.~(\ref{cartanobs}), defining the components (\ref{split1}) with respect to the observer field $u$ or its internal counterpart defined by $y=e(u)$.

After a choice of observer field, we are hence left with an $\Iso(D,1)$-valued 1-form on $\M$ for which the $\so(D)$ part is an Ehresmann connection on $Q_u\to\M$ and for which the projection to $\so(D)\oplus\R^{D}\oplus\R$ (projecting out boosts) is a Cartan connection on $Q_u\to\M$, which models spacetime on $(\SO(D)\ltimes\R^{D+1})/\SO(D)$, an absolute-time version of Minkowski spacetime. For more details, see Ref.~\cite{obspaper}.

Observer space allows us a perspective on canonical gravity {\em \`a la} Ashtekar where all choices of foliation can be considered simultaneously and hence provides a link between the canonical and covariant approaches to general relativity \cite{essay}. It directly links the space-time split induced by a choice of foliation of spacetime with a splitting of representations in the ``internal'' gauge group $\SO(D,1)$. It replaces the lapse and shift of Arnowitt--Deser--Misner (ADM) variables with a field that has a more direct interpretation, and that can play a more fundamental role in gravitational physics; for instance, ``energy density'' is simply a scalar on observer space.

It also allows us to recover Lorentz covariance from the $\SU(2)$ Ashtekar--Barbero formulation, for instance, but of course we knew already that the original theory defined by Eq.~(\ref{holstak}) was Lorentz covariant, describable purely in terms of the components $(\omega,e)$ of an $\ISO(D,1)$ Cartan connection on $\M$. We started off with the orthonormal frame bundle $\fb\to \M$, which implements local Lorentz symmetry and where all the dynamical fields live.

In this paper we take a slightly different viewpoint: we assume a gravitational theory given in terms of fields on an $\SO(D)$ bundle $Q\to\M$. We then aim to reconstruct the associated observer space $P\to O$ by first exhibiting the choice of observer field implicit in the original choice of variables and then finding the transformations relating this observer field to others. These transformations enlarge the group $\SO(D)$ to some larger group $H'$; the fibers of the reconstructed bundle $O\to\M$ have the structure of homogeneous spaces $H'/\SO(D)$, and there is a projection $P\to\M$ making $P$ into a principal $H'$ bundle over $\M$.

The possibilities for what $H'$ can be are limited; in their seminal paper \cite{bacryll}, Bacry and L\'evy-Leblond classified all {\em kinematical groups} only assuming a rotational subgroup acting on ``boosts,'' ``space translations'' and ``time translations,'' and the existence of parity and time-reversal as automorphisms on the kinematical group.\footnote{While Ref.~\cite{bacryll} works with $D=3$, their derivation seems to be applicable to arbitrary $D$.} In this paper we work with a flat model spacetime so that space and time translations commute; since we also work in a preferred foliation framework, it is natural to take $H'$ as one of the {\em absolute-time groups} in Ref.~\cite{bacryll}. These two assumptions restrict us to either the Galilei group $\ISO(D)\ltimes\R^{D+1}$ or the static group $\SO(D)\ltimes \R^{2D+1}$.

As a warmup exercise, let us start off by looking at the observer space associated to spacetime in the well-known geometric formulation of Newtonian mechanics given by Cartan.

\sect{Newton-Cartan Gravity}
\label{newtsec}

In Newtonian mechanics, spacetime $\M$ is of the form $\M\cong \S \times \R$ for some $D$-dimensional manifold $\S$ of fixed geometry, not necessarily Euclidean, so that we generalize Newtonian mechanics by allowing a curved background space. Spacetime comes with a closed 1-form $\tau\equiv dt$, which is orthogonal to the constant-time hypersurfaces $\S_t$, and with a degenerate inverse metric
\ben
{\bf h} = {\bf \vec{e}}_i \otimes {\bf \vec{e}}_j \,\delta^{ij}
\label{newtmet}
\een
for some triad of vectors ${\bf \vec{e}}_i$, which is spatial, $\tau({\bf \vec{e}}^i)=0$; hence, ${\bf h}(\tau,\tau)=0$. 

In order to define the inverse of Eq.~(\ref{newtmet}), or alternatively a cotriad $\vec{e}^i$, one needs to have a notion of ``spatial'' for covectors. For this it is necessary to fix a field of observers, a vector field $u$ on $\M$ such that $\tau(u)=1$. This vector field is not unique (see the discussion in Ref.~\cite{newtcart}, Sec. 5); the set of possible observer fields is related to any given observer field $u$ by
\ben
u' = u + v^i\,{\bf \vec{e}}_i
\label{obtr}
\een
where $v^i$ is an arbitrary $\R^D$-valued function on $M$.

After a choice of observer field $u$, the metric (\ref{newtmet}) defines an isomorphism between the subspaces of spatial covectors (covectors annihilating $u$) and spatial vectors; on these subspaces its inverse can be written as
\ben
 h = \vec{e}^i \otimes \vec{e}^j\, \delta_{ij}
\een
where $\vec{e}^i$ is a basis in the space of spatial covectors dual to the spatial triad, $\vec{e}^i({\bf \vec{e}}_j)=\delta^i_j$. Notice the close analogy to the construction for Ashtekar--Barbero variables given in Ref.~\cite{asht} and outlined in Sec. \ref{cartansec}, where a foliation is specified by a 1-form $\hat{u}$, and the dynamical variable $y$, an internal field of observers, gives a notion of spatial and temporal differential forms through its associated spacetime observer field defined by $y=e(u)$.

A spin connection encodes the Newtonian dynamics of point particles: its $\so(D)$ part is $\Gamma[\vec{e}]$, the torsion-free (Levi--Civita) connection compatible with the cotriad $\vec{e}^i$, while its $\R^D$ part encodes the acceleration due to a Newtonian force,
\ben
{\omega^i}_0 = -F^i \,\tau\,.
\label{force}
\een

Observer space $O$ should be a bundle over spacetime $\M$, for which the fibers are vector spaces $\R^D$ relating the different observers by Eq.~(\ref{obtr}). To define a Cartan geometry over $O$, we need to choose a Lie group $G$ which has $\SO(D)$ as a subgroup so that $G/SO(D)$ can serve as a model observer spacetime. As explained above $G$ should be a kinematical group in the sense of Ref.~\cite{bacryll} which preserves absolute time and which includes commuting spacetime translations. 

To obtain nontrivial gauge transformations that could relate different observers according to Eq.~(\ref{obtr}), let us take $G$ to be the inhomogeneous Galilei group $\ISO(D)\ltimes\R^{D+1}$. Then we must first pick a particular observer field $u$ and define a 1-form on $\M$ [really the pullback of a 1-form on the $\SO(D)$ bundle $Q\to \M$ by some section] valued in the Lie algebra $\Iso(D)\oplus\R^{D+1}$ that is to become a Cartan connection on $O$,
\ben
A = \begin{pmatrix} \Gamma[\vec{e}] & -F & \vec{e}/l \\ 0 & 0 & \tau/l \\ 0 & 0 & 0 \end{pmatrix}\,,
\label{cartanconn}
\een
where we should, if we worry about physical dimensions, rescale $\vec{e}$ and $\tau$ with a length scale $l$. We want to interpret $\Gamma$ and $F$ as parts of an $\ISO(D)$ spin connection on $\M$.

The curvature of $A$ is
\ben
F[A] = dA+\half[A,A]=\begin{pmatrix} R[\vec{e}] & -d_{\Gamma}F & 0 \\ 0 & 0 & 0 \\ 0 & 0 & 0 \end{pmatrix}
\label{newcurv}
\een
where we assume that the cotriad $\vec{e}$ is constant in time, in the sense that
\ben
d\vec{e}(u)=0\,,
\label{constant}
\een
so that apparently there is no ``evolution of space'' in Newtonian gravity. Note that only the spatial derivatives of the Newtonian force $F$ contribute to Eq.~(\ref{newcurv}).

While $\SO(D)$ transformations now just have the usual effect of local frame rotations, a nonrelativistic boost corresponds to the transformation
\ben
\vec{e}\mapsto\vec{e}-b\,\tau\,,\quad -F\mapsto -F+\Gamma\cdot b+db
\label{obstrafo}
\een
for some $\R^D$-valued function $b$ on $\M$. Such a transformation changes the notion of observer: the transformed cotriad $\vec{e}'=\vec{e}-b\,\tau$ annihilates a different observer field, namely,
\ben
\left(\vec{e}-b\,\tau\right)\left(u+b\cdot{\bf e}\right) = b - b = 0\,,
\een
which is consistent with Eq.~(\ref{obtr}). Hence, it seems that Newton--Cartan gravity is described by the 1-form on $O$,
\ben
A = \begin{pmatrix} \Gamma[\vec{e}] & -F + \Gamma\cdot b + db & \vec{e}-b\,\tau \\ 0 & 0 & \tau \\ 0 & 0 & 0 \end{pmatrix}\,,
\label{cartanconn2}
\een
where $\Gamma,\vec{e}$ and $\tau$ are constant along the fibers of $O\to M$ and $b$ are coordinates on these fibers. $A$ can be viewed as the pullback of a connection on an $\SO(D)$ bundle $P\to O$ to $O$, defining a Cartan geometry on $O$ modelled on the nonrelativistic observer space $(\ISO(D)\ltimes\R^{D+1})/\SO(D)$. However, there is a constraint: for the transformed boost part $F'$ to be still of the form (\ref{force}), the functions $b$ must be spatially covariantly constant,
\ben
\left(d b^i + {\Gamma^i}_j\,b^j\right)({\bf \vec{e}}_k) = 0\quad\forall k\,.
\label{covcons}
\een
The same condition is obtained by requiring that $\Gamma[\vec{e}]$ is compatible with the transformed cotriad $\vec{e}'=\vec{e}-b\,\tau$ as well. The part of $db$ proportional to $\tau$ is then an ``accelerative force'' modifying the Newtonian equations of motion; such transformations are usually not regarded as transformations between inertial frames.

Using the projection $O\to \M$, we could view $A$ as defining a connection on the $\ISO(D)$ bundle $P\to\M$ together with an $\R^{D+1}$-valued 1-form defining a coframe field. However, Newton--Cartan gravity is {\em not} a gauge theory on this nonrelativistic frame bundle; there is a preferred class of observers, related by transformations satisfying Eq.~(\ref{covcons}), that preserve the structure of the Newtonian equations of motion as geodesic equations on spacetime $\M$ and the compatibility of the connection $\Gamma$ with the co-triad $\vec{e}$. Physics takes place on an $\SO(D)$ bundle defined for any of the preferred observers. The time independence of spatial geometry (\ref{constant}) is not preserved under general observer transformations either, but it is satisfied for a Euclidean cotriad constant over $\M$ so that $d\vec{e}=0$.

The ``coframe field'' formed by $(\vec{e},\tau)$ defines an isomorphism between $\R^{D+1}$ and the tangent space $T_x\M$ at each point in $\M$, but since there is no appropriate nondegenerate symmetric form on $\R^{D+1}$ invariant under $\ISO(D)$, one cannot construct a nondegenerate metric out of it. We can define a metric on spacetime
\ben
\hat{g} = \vec{e}^i \otimes \vec{e}^j\, \delta_{ij} \pm \tau \otimes \tau\,,
\een
where one has a choice of relative sign. This ``spacetime metric'' depends on the observer field $u$ and is therefore not an intrinsic geometric structure; it is not invariant under $\ISO(D)$.

Taking as kinematical group $G$ the static group $\SO(D)\ltimes\R^{2D+1}$ instead of the inhomogeneous Galilei group $\ISO(D)\ltimes \R^{D+1}$ trivializes the transformation of the cotriad $\vec{e}$ but not the action of boosts on the Newtonian force,
\ben
\vec{e}\mapsto\vec{e}\,,\quad -F\mapsto -F+\Gamma\cdot b+db\,,
\een
so that this does not solve the issue.

Formally, there seems to be an alternative way to construct a ``nonrelativistic frame bundle'' over $\M$: there is another $\ISO(D)$ subgroup of $\ISO(D)\ltimes\R^{D+1}$ consisting of spatial rotations and spatial translations. The latter leave the Newtonian force invariant:
\ben
\vec{e}\mapsto\vec{e}+\Gamma\cdot t+dt\,,\quad -F\mapsto -F\,.
\een
Generic transformations of this form can again be interpreted as a shift in observer and seem to leave invariant the equations of Newtonian mechanics. Those, however, no longer arise as geodesic equations associated to the ``spin connection,'' which now consists of $(\Gamma,\vec{e})$ [the spacetime version of the connection arising in Cartan geometrodynamics (\ref{cartangeo})]. Worse still, the components of $F$ are all proportional to the 1-form $\tau$, and hence the $\R^D$-valued 1-form $F$ does not define a nondegenerate cotriad on space. The corresponding 1-form $A$ cannot be interpreted as a Cartan connection on the ``alternative nonrelativistic frame bundle'' $P'\to \M$.

\sect{Ho\v{r}ava-Lifshitz Gravity}

Let us apply the insights gained from Newton--Cartan gravity to a gravitational theory of dynamical spacetime with a preferred foliation. We mainly think of the theory proposed by Ho\v{r}ava \cite{horava}, a metric theory with a preferred foliation, where one is supposed to think of space and time as coming with different physical dimensions, i.e., different conformal weights under the action of a dilatation; but our constructions will be general and applicable to any framework involving a foliation of spacetime.

Just as in Sec. \ref{newtsec}, spacetime $\M$ comes with a 1-form $\tau$, assumed to be closed, determining a foliation of spacetime by spatial hypersurfaces $\M\cong\S\times\R$.\footnote{The requirement on a 1-form $\sigma$ to be hypersurface-orthogonal is $\sigma\wedge d\sigma=0$; any such $\sigma$ is of the form $\sigma=N\,\tau$ where $\tau$ is closed (and so at least locally a gradient) and $N$ is a spacetime function called the lapse.} The basic variables are a three-metric $h$ on spatial slices and the extrinsic curvature ${\bf K}$, together with a shift vector $\vec{N}$ and a lapse function $N$. The basic symmetries of the theory are foliation-preserving diffeomorphisms:
\ben
x \mapsto x' = x'(x,t)\,,\quad t \mapsto t' = t'(t)\,.
\een
These look like a local version of inhomogeneous Galilei transformations; in Ref.~\cite{horava1} Ho\v{r}ava derives the action on the fields of the theory by considering the nonrelativistic limit $c\rightarrow\infty$ of usual diffeomorphisms in the ADM formalism. It is precisely the same limit that gives an In\"on\"u--Wigner contraction \cite{inonu} of the Lorentz group to the Galilei group, which hints at an alternative formulation of Ho\v{r}ava--Lifshitz gravity as a theory on a principal $\ISO(D)$ bundle.

We shall make this relation more precise by rephrasing Ho\v{r}ava--Lifshitz gravity in connection form. In the philosophy of Ho\v{r}ava--Lifshitz gravity, we must associate physical dimensions to space and time coordinates so that
\ben
[dx^{\alpha}]=-1\,,\quad[\tau]=[dt]=-z
\een
where we define dimensions with respect to spatial momenta, or inverse spatial lengths.

In a first-order Palatini formulation, the dynamical variables of gravity are a spatial cotriad $\vec{e}$, an $\mathfrak{so}(D)$ connection $\Omega$, and an $\R^D$-valued 1-form $K$ related to the extrinsic curvature, for which the spatial components are the usual canonical variables for gravity defined on a given foliation. They come with momentum dimensions\footnote{The components of $\vec{e}$ in a coordinate basis would be dimensionless so that $\vec{e}=\vec{e}_{\alpha}\,dx^{\alpha}$ can have dimension $-1$.}:
\ben
[\vec{e}]=-1\,,\quad [\Omega]=0\,,\quad [K]=z-1\,.
\een

In the usual ADM formulation \cite{adm}, one would define a spacetime metric by
\ben
g = -N^2 \tau\otimes\tau + \left(\vec{e}^i + \vec{N}^i\tau\right)\otimes\left(\vec{e}^j + \vec{N}^j\tau\right)\delta_{ij}\,,
\label{stm}
\een
where $\vec{N}^i$ is related to the spatial vector field $N^{\alpha}$ usually defined as the shift vector by $\vec{N}^i=\vec{e}^i(N)$. However, in the formalism of Ho\v{r}ava--Lifshitz gravity, the more natural type of metric on $\M$ is a spatial metric built out of the cotriad $\vec{e}$ and a lapse $\vec{N}$,
\ben
 h = \left(\vec{e}^i + \vec{N}^i\tau\right)\otimes\left(\vec{e}^j + \vec{N}^j\tau\right)\delta_{ij}\,,
\label{3m}
\een
since in Refs.~\cite{horava,horava1} the lapse is dimensionless so that neither $\tau$ nor $N\tau$ have the right dimension. Note that with respect to the metric (\ref{3m}), the orthonormal cotriad is $\vec{e}+\vec{N}\,\tau$ and not $\vec{e}$. Consistency requires $\vec{N}$ to have dimension $z-1$.

The metric (\ref{3m}) is invariant under local $\SO(D)$ rotations acting on $\vec{e}$ and $\vec{N}$; there is also an additional shift invariance under
\ben
\vec{e}\mapsto \vec{e} - \vec{M}\,\tau\,,\quad \vec{N}\mapsto \vec{N}+\vec{M}
\label{shiftinv}
\een
for any $\R^D$-valued scalar $\vec{M}$ (of momentum dimension $z-1$). This additional invariance (which is not there in the metric formulation where there is no cotriad) provides another hint at the possibility of extending the local gauge symmetry from $\SO(D)$ to $\ISO(D)$, which we will now investigate.

Just as we tried for Newton--Cartan gravity, we can use the dynamical fields to build a 1-form $A$ on an $\SO(D)$ bundle $Q$ over $\M$ that will supposedly become a Cartan connection on some observer space bundle $P\to O$ yet to be defined. Again we have a choice of Lie group $G$, which we take to be the inhomogeneous Galilei group, and consider the following $\Iso(D)\oplus\R^{D+1}$-valued 1-form defined on spacetime $\M$:
\ben
A = \begin{pmatrix} \Omega & l^{z-1}\,K & \vec{e}/l \\ 0 & 0 & N\,\tau/l^{z} \\ 0 & 0 & 0 \end{pmatrix}\,.
\label{cartanhor}
\een
Since $A$ is to be dimensionless, we have rescaled $\vec{e},\tau,$ and $K$ accordingly; we are also using the normalized 1-form $N\,\tau$ instead of the closed 1-form $\tau$ so that $(\vec{e},N\,\tau)$ can be interpreted as an orthonormal cotriad. It is rather unclear what sets the spatial length scale $l$.\footnote{In so-called ``Lifshitz spacetimes,'' which have {\em global isometries} interpretable as implementing anisotropic scaling between space and time directions, such a scale is associated to the (negative) cosmological constant. For a recent discussion of the possible interpretation of such spacetimes as solutions of Ho\v{r}ava--Lifshitz gravity, see Ref.~\cite{lifshitz}.} Note, however, that the speed of light, as any velocity, now has momentum dimension $z-1$ so that we may take
\ben
A = \begin{pmatrix} \Omega & K/c & \vec{e}/l \\ 0 & 0 & c\,N\,\tau/l \\ 0 & 0 & 0 \end{pmatrix}
\label{horconn}
\een
and the scale $l$ is now arbitrary just as in the formulation of gravity as a Poincar\'e gauge theory.

Computing the curvature of $A$, we find
\ben
F[A] = \begin{pmatrix} R[\Omega] & d_{\Omega}K/c & \left(d_{\Omega}\vec{e}+N\,K\wedge\tau\right)/l \\ 0 & 0 & c\,d(N\,\tau)/l \\ 0 & 0 & 0 \end{pmatrix}\,.
\label{galilcur}
\een

We now want to lift the 1-form $A$ on $\M$, defined with respect to a particular ``observer,'' to a 1-form on $O$ so that $O$ can become a Cartan geometry modelled on $(\ISO(D)\ltimes\R^{D+1})/\SO(D)$. Hence, we must understand the analog of Newtonian observer transformations (\ref{obstrafo}).

 The effect of Galilean boosts on the cotriad is precisely the one anticipated in Eq.~(\ref{shiftinv}); the boost part of $A$ is transformed as well:
\ben
\vec{e}\mapsto\vec{e}-b\,N\,\tau\,,\quad K\mapsto K+\Omega\cdot b+db \equiv K + d_{\Omega}b\,.
\label{gali}
\een
Identifying the first of those transformations with Eq. (\ref{shiftinv}), the second one may be written as
\ben
K\mapsto K + d_{\Omega}\left(\delta\vec{N}/N\right)\,.
\een

To understand this transformation behavior of $K$, we need to clarify the relationship between the 1-form $K$ appearing in Eq.~(\ref{horconn}) and the usual boost part of the curvature ${\bf K}$ for which the spatial components define the extrinsic curvature. From the spacetime perspective, given a cotriad $(\vec{e},N\,\tau)$, one can uniquely determine an $\SO(D,1)$ spin connection $(\Omega,{\bf K})$ by looking at Cartan's equations of structure requiring torsion to vanish,
\bena
d\vec{e} + [\Omega,\vec{e}] + N\,{\bf K}\wedge\tau &=& 0\,,
\label{cartan1}
\\ d(N\,\tau) + {\bf K}\wedge\vec{e} &=& 0\,.
\label{cartan3}
\eena
The spatial part of this ${\bf K}$ is the extrinsic curvature of the spacetime geometry specified by $(\vec{e},\tau)$ if we take $\tau$ as normal to spatial hypersurfaces. 

Equations (\ref{cartan1}) and (\ref{cartan3}) arise as equations of motion from the Einstein--Hilbert--Palatini action (\ref{akshn}). They are equivalent to the vanishing of the translational part of the curvature of the usual Cartan connection, valued in $\Iso(D,1)$ for a vanishing cosmological constant.

In our setting where observer space is nonrelativistic, it is natural to require that the translational part of Eq.~(\ref{galilcur}) vanishes, leading to the Galilean Cartan equations of structure:
\bena
d\vec{e} + [\Omega,\vec{e}] + N\, K\wedge\tau &=& 0\,,
\label{cartan4}
\\d(N\,\tau) &= &0\,.
\label{cartan2}
\eena
The transformation of $K$ in Eq.~(\ref{gali}) preserves Eqs.~(\ref{cartan4}) and (\ref{cartan2}) under a shift in the cotriad $\vec{e}$.

Equation (\ref{cartan2}) is no longer an equation that determines components of the connection but a condition on the coframe instead; it requires the lapse to be a function of time only, $N=N(t)$. The same condition arises in the original ``projectable'' version of Ho\v{r}ava--Lifshitz gravity in Ref.~\cite{horava} from a different viewpoint; here it is simply equivalent to setting part of the curvature of $A$ to zero. 

For closed $\tau$ and projectable lapse, any connection $(\Omega,{\bf K})$ that solves Eqs.~(\ref{cartan1}) and (\ref{cartan3}) also solves Eqs.~(\ref{cartan4}) and (\ref{cartan2}), but the converse is not true. To understand more precisely how the equations of structure constrain the connection, we can adopt a canonical viewpoint and split Eqs.~(\ref{cartan1}) and (\ref{cartan3}) further: their projections to the span of $\vec{e}\wedge\vec{e}$ are {\em constraint equations},
\bena
\nabla\vec{e} + [\Omega_{\vec{e}},\vec{e}] &=& 0\,,
\label{levic}
\\ {\bf K}_{\vec{e}}\wedge\vec{e} &=& 0\,.
\label{gauss}
\eena
Here we denote by $X_{\vec{e}}$ the projection of a 1-form to the subspace of spatial covectors spanned by $\vec{e}$; similarly we have split $d\vec{e}$ as $d\vec{e}=\nabla\vec{e}+\tau\wedge\partial_t\vec{e}$. We also continue to assume that $N\,\tau$ is hypersurface-orthogonal so that $d(N\,\tau)=\tau\wedge\sigma[\tau]$ for some $\sigma$.

Equation (\ref{gauss}) is the usual ``Gauss constraint'' of first-order canonical gravity (see, e.g., Ref.~\cite{thiemann}), while Eq.~(\ref{levic}) fully determines $\Omega_{\vec{e}}$. The remaining pieces of Eqs.~(\ref{cartan1}) and (\ref{cartan3}) are evolution equations,
\bena
\tau\wedge \left(\partial_t\vec{e}+\Omega_\tau\cdot\vec{e}-N\,{\bf K}_{\vec{e}}\right) &=& 0\,,
\label{evo1}
\\\tau\wedge\left(\sigma[\tau]+{\bf K}_\tau\vec{e}\right) &=& 0\,.
\label{evo2}
\eena
The first of those can be solved for $\Omega_{\tau}$ and ${\bf K}_{\vec{e}}$: expanding the $\R^D$-valued 1-forms appearing in Eq.~(\ref{evo1}) in the basis $\vec{e}$, we get
\ben
(\partial_t\vec{e})^{ij}+\Omega_\tau^{ij}-N\,{\bf K}_{\vec{e}}^{ij}=0\,,
\een
which fixes $-\Omega_{\tau}$ as the antisymmetric and $N\,{\bf K}_{\vec{e}}$ as the symmetric part of $\partial_t\vec{e}$; the extrinsic curvature ${\bf K}_{\vec{e}}$ must be symmetric by Eq.~(\ref{gauss}). Equation (\ref{evo2}) then fixes the remaining components ${\bf K}_\tau$ of $(\Omega,{\bf K})$.

Now consider a 1-form $K$ solving just Eq.~(\ref{cartan4}). There is no constraint (\ref{gauss}) on $K$ to be symmetric; there is also no Eq.~(\ref{evo2}) to fix $K_\tau$. It follows that the general solution to Eq.~(\ref{cartan4}) is
\ben
K = {\bf K} + \chi + \Xi\,\tau\,,
\label{ksplit}
\een
where $\chi$ is in the span of $\vec{e}$ and antisymmetric: $\chi^{(ij)}=0$, where $\chi^i\equiv\chi^{ij}\vec{e}_j$.

We can use the freedom to choose $\chi$ and $\Xi$ while preserving Eq.~(\ref{cartan4}) to obtain a 1-form $K$ that transforms under changes in the shift vector as
\ben
K\mapsto K + d_{\Omega}\left(\delta\vec{N}/N\right)\,.
\label{ktrafo}
\een
This is straightforward. First note that consistency of Eq.~(\ref{ktrafo}) with Eq.~(\ref{ksplit}) requires that
\ben
{\bf K}^{ij}\mapsto {\bf K}^{ij} + \frac{1}{2N}\left(\nabla^i \delta N^j + {\Omega^{ij}}_k \delta N^k +(i\leftrightarrow j)\right)\,,
\een
using the projectability of the lapse. This is precisely the transformation of the extrinsic curvature under a change in the shift vector; we use the conventions of Ref.~\cite{wilthsire} \mbox{for ${\bf K}$,}
\ben
{\bf K}_{\alpha\beta} = \frac{1}{2N}\left(N_{\alpha;\beta}+N_{\beta;\alpha}-\frac{\partial h_{\alpha\beta}}{\partial t}\right)\,,
\een
in a coordinate basis. The fields $\chi$ and $\Xi$ can then be chosen to be
\bena
\chi^{ij} & = & -\frac{1}{2N}\left(\nabla^i N^j + {\Omega^{ij}}_k N^k -(i\leftrightarrow j)\right)\,,\nonumber
\\\Xi^i & = & \partial_t \left(N^i/N\right) + {{\Omega_{\tau}}^i}_j N^j/N\,.
\eena

Having thus established a consistent action of Galilean boosts on an $\Iso(D)\oplus\R^{D+1}$-valued 1-form, this 1-form can be lifted to a Cartan connection on the nonrelativistic frame bundle as an $\SO(D)$ bundle over observer space $O$:
\begin{widetext}
\ben
A(x,\vec{N}) = \begin{pmatrix} \Omega(x) & \left({\bf K}_0(x)+d_{\Omega}(\frac{\vec{N}}{N(t)})\right)/c & \left(\vec{e}(x)-\vec{N}\tau\right)/l \\ 0 & 0 & c\,N(t)\,\tau/l \\ 0 & 0 & 0 \end{pmatrix}\,.
\label{supercartan}
\een
\end{widetext}
Observer space is a vector bundle over spacetime $\M$ with fibers isomorphic to $\R^D$, for which the geometry is described by the Cartan connection $A$ comparing it infinitesimally to a model observer space $(\ISO(D)\ltimes\R^{D+1})/\SO(D)$. We have picked a trivialization of $O\to\M$ and coordinates $\{x\}$ on $\M$ and $\vec{N}$ on the fibers to make the structure of the Cartan connection explicit.\footnote{This means ``geometrizing'' the shift vector from a spacetime field to defining coordinates in a fiber of $O\to\M$.} ${\bf K}_0$ is the boost part of the normal torsion-free spin connection, associated to zero shift vector, i.e., the section of $O\to\M$ given by $\vec{N}=0$ in our coordinates.

As a consistency check, one can compute the curvature of $A$, which is
\ben
F[A] = \begin{pmatrix} R[\Omega] & \left(d_{\Omega}{\bf K}_0+R[\Omega]\cdot \frac{\vec{N}}{N}\right)/c & 0 \\ 0 & 0 & 0 \\ 0 & 0 & 0 \end{pmatrix}\,.
\een
Here $R[\Omega]=d\Omega+\half[\Omega,\Omega]$ is the curvature of $\Omega$, constant along the fibers since $\Omega$ already is. The 2-form $F$ annihilates any vector tangent to the fibers, as \mbox{it should be.}

The connection (\ref{supercartan}) on $O$ becomes a Cartan connection on the nonrelativistic frame bundle $P\to\M$, which models a Ho\v{r}ava--Lifshitz spacetime $\M$ on nonrelativistic spacetime $(\ISO(D)\ltimes\R^{D+1})/\ISO(D)$. A choice of section in this bundle amounts to picking a local spatial frame, together with a shift vector $\vec{N}$. The spatial metric is then not the one formed by ``squaring'' the spatial cotriad $\vec{e}$ appearing in the translational part of the Cartan connection; it is the linear combination of this cotriad and the shift,
\ben
 h = \left(\vec{e}^i + \vec{N}^i\tau\right)\otimes\left(\vec{e}^j + \vec{N}^j\tau\right)\delta_{ij}\,.
\een
It is only this metric that can be $\ISO(D)$-invariant, as we have anticipated above.

To extract the physical variables of Ho\v{r}ava--Lifshitz gravity from a given torsion-free Cartan connection on the bundle $P\to\M$, one can pick a section of the bundle to obtain an $\Iso(D)\oplus\R^{D}\oplus \R$-valued 1-form on $\M$. Its $\R$ part gives the (normalized) normal to the foliation while its $\R^D$ part can be combined with the shift vector determined by the section to give a spatial metric $h$; the projection of its boost part to the span of the cotriad then includes the extrinsic curvature ${\bf K}_{\vec{e}}$ as its symmetric part. This projection is not $\ISO(D)$-invariant.

The absence of a suitable nondegenerate tensor invariant under $\ISO(D)$ makes it difficult to construct explicitly $\ISO(D)$-invariant theories, in much the same way that the MacDowell--Mansouri formulation of gravity only works for nonzero cosmological constant where the gauge group is $\SO(D,2)$ or $\SO(D+1,1)$ instead of the Poincar\'e group $\ISO(D,1)$.

\sect{Conclusions}
There are many ways of viewing gravity as a gauge theory, with a connection on a principal fiber bundle as its dynamical variable. A compelling geometric picture encompassing many such approaches is provided by Cartan geometry, where pure gravity is essentially described by a single dynamical field: a Cartan connection valued in the Lie algebra $\g$ of a group $G$ on a fiber bundle with structure group $H\subseteq G$, obtained by reducing a principal $G$ fiber bundle by a symmetry-breaking field valued in $G/H$. The symmetry-breaking field can be viewed as corresponding to a choice of base point in a ``tangent spacetime'' in covariant gravity or a choice of observer in canonical gravity, where a foliation determines an observer at each point. 

Ashtekar--Barbero variables define a Cartan geometry which involves choosing both a point and an observer in a model spacetime and is best understood on observer space. Since Ashtekar--Barbero variables can be derived from the Lorentz covariant Palatini formulation, the Cartan connection becomes an $\Iso(D,1)$-valued 1-form on a principal $\SO(D)$ bundle over an observer space that is modelled on the homogeneous space $\ISO(D,1)/\SO(D)$ of observers in Minkowski spacetime.

In this paper, we have investigated nonrelativistic observer space geometries, where different observers are related not by Lorentz but by Galilei transformations. Starting from Newton--Cartan gravity where only certain preferred observers are allowed, we constructed the nonrelativistic observer space of gravity with a fixed foliation. Local Galilean boosts preserve the foliation while changing the shift vector of canonical gravity, which plays the role of a symmetry-breaking field valued in $\R^D\cong\ISO(D)/\SO(D)$. Including the shift vector into the definition of the Cartan connection allows the restoration of full Galilei invariance. 
The $\SO(D)$ bundle over nonrelativistic observer space becomes the {\em nonrelativistic frame bundle} with structure group $\ISO(D)$ over spacetime; picking a section in this bundle, the Cartan connection can be viewed as a 1-form including the dynamical fields of gravity with respect to a particular choice of shift vector.

We have not discussed the dynamics of gravity in any detail, but kinematically gravity does not have to be a theory with either local Lorentz invariance or invariance under a rotational subgroup; it can equally well be a theory with local Galilei invariance, where Lorentz invariance is not broken but {\em contracted}. This description is adapted to preferred foliation theories and the gauge-theoretic analog of the foliation-preserving transformations considered in the metric framework of Ref.~\cite{horava}.

Of course, for pure gravity the discussion about these different descriptions is somewhat academic; only when matter is coupled can different gravitational gauge groups lead to different physics. We leave a discussion of matter coupling, as well as a more detailed application of the geometric picture presented here to various preferred foliation theories, to future work.

\subsection*{Acknowledgements}
I would like to thank Derek Wise for useful suggestions on the manuscript. I am supported by a Riemann Fellowship of the Riemann Center for Geometry and Physics. Research at Perimeter Institute is supported by the Government of Canada through Industry Canada and by the Province of Ontario through the Ministry of Research and Innovation.

\end{document}